\documentclass[10pt]{article}
\usepackage{authblk}
\usepackage{graphicx}
\usepackage{epstopdf}

\begin{document}

\title{Solution of the Master Equation for Quantum Brownian Motion given by
the Schr\"{o}dinger Equation}
\author[1]{R. Sinuvasan}
\author[2,3]{A Paliathanasis\thanks{%
anpaliat@phys.uoa.gr}}
\author[4]{R.M Morris}
\author[4,5]{ P.G.L. Leach\thanks{%
leach@ucy.ac.cy}}

\affil[1]{Department of Mathematics, Pondicherry University, Kalapet
Puducherry 605 014, India}
\affil[2]{Instituto de Ciencias F\'{\i}sicas y Matem\'{a}ticas, Universidad Austral de
Chile, Valdivia, Chile}
\affil[3]{Institute of Systems Science, Durban
University of Technology, PO Box 1334, Durban 4000, Republic of South Africa}
\affil[4]{Department of Mathematics and
Institute of Systems Science, Research and Postgraduate Support, Durban
University of Technology, PO Box 1334, Durban 4000, Republic of South Africa}
\affil[5]{School of Mathematics, Statistics and Computer Science, University
of KwaZulu-Natal, Private Bag X54001, Durban 4000, Republic of South Africa}

\renewcommand\Authands{ and }
\maketitle

\begin{abstract}
We consider the master equation of quantum Brownian motion and with the
application of the group invariant transformation we show that there exists
a surface on which the solution of the master equation is given by an
autonomous one-dimensional Schr\"{o}dinger Equation.
\end{abstract}


\section{Introduction}

With the method of path integrals, and specifically the Feynman-Vernon
influence functional \cite{Feynman}, Haake and Reibold \cite{haake} and few
years latter Hu, Paz and Zhang derived an equation which inherits the
properties of quantum Brownian motion for a harmonic oscillator interacting
with a linear passive heat bath of oscillators \cite{hu1,hu2}. An alternate
derivation of that master equation has been performed by Halliwell and Yu by
tracing the evolution equation for the Wigner function of the system \cite%
{halli}.

The master equation of quantum Brownian motion is an $\left( 1+2\right) $
linear nonautonomous evolution equation given by
\begin{equation}
Z_{,t}=-\frac{x}{m}Z_{,y}+m\Omega ^{2}(t)yZ_{,x}+2\Gamma (t)(xZ)_{,x}+%
\hbar%
m\Gamma (t)h(t)Z_{,xx}+%
\hbar%
\Gamma (t)f(t)Z_{,xy},  \label{eq.01}
\end{equation}%
where $m$ is the mass of the Brownian particle, $Z=Z\left( t,x,y\right) $ is
the Wigner function of the density matrix ($x$ denotes the momentum of the
oscillator and $y$ its position). Furthermore, the coefficients, $\Omega
^{2}\left( t\right) $,~$\Gamma \left( t\right) ,~h\left( t\right) $ and $%
f\left( t\right) $, in general are time-dependent and related to the natural
frequency of the Brownian motion and the terms interacting with the heat
bath of oscillators. The derivation of the coefficients is given in \cite%
{halli,ford}.

A general analytical solution of the master equation (\ref{eq.01}) with the
use of the Langevin Equation has been derived in \cite{ford}, whereas in
\cite{solmaster} some solutions of the master equations for quantum Brownian
motions are given. The analysis of open quantum system does not stop in
equation (\ref{eq.01}). A relation of the exact master equation with the
nonequilibrium Green functions for non-Markovian open quantum systems was
derived in \cite{WM1}, while new phenomena concerning the thermal-state of a
quantum system were predicted for a strongly non-Markovian enviroment in
\cite{WM2}.

In this work we are interested in the existence of solutions for equation (%
\ref{eq.01}) which follow from the method of group invariant
transformations, in particular we are interested in the one-parameter point
transformations which were introduced by S. Lie \cite{lie}, where the
generator of the infinitesimal transformation is called a Lie (point)
symmetry. The importance of Lie symmetries is that they provide a systematic
method to facilitate the solution of differential equations because they
provide first-order invariants which can be used to reduce the order of
differential equations. Moreover Lie symmetries can be used for the
classification of differential equations and important information for the
differential equation can be extracted from the admitted group of invariant
transformations. The method of group invariant transformations has been
applied in various systems of quantum mechanics, for instance see \cite%
{qm1,qm2,qm3,qm4,qm5,qm6} and references therein.

By applying the Lie theory for differential equations we show that equation (%
\ref{eq.01}) is invariant under a group of one-parameter point
transformations in which the generators form the \ $\{A_{1}\oplus
_{s}W_{5}\}\oplus _{s}\infty A_{1},~$Lie algebra, where $W_{5}$ denotes the
five-element Weyl-Heisenberg algebra and $\infty A_{1}$ is the
infinite-dimensional abelian algebra of the solutions of the linear $(1+2)$
evolution equation and follows from the linearity of (\ref{eq.01}).
Furthermore from the Lie symmetries we can define a surface in which
equation (\ref{eq.01}) is independent of one of the independent variables
and with the use of the zeroth-order invariants we can reduce equation (\ref%
{eq.01}) in a nonautonomous one-dimensional evolution equation. We study the
Lie point symmetries of this equation and show that is maximally symmetric.
Hence it is invariant under a group of transformation which form the%
\footnote{%
Algebra $sl(2,R)$ is the $A_{3,8}$ and $W_{3}$ is the $A_{3,3}$ in the
Mubarakzyanov Classification Scheme \cite%
{Morozov58a,Mubarakzyanov63a,Mubarakzyanov63b,Mubarakzyanov63c}} $%
\{sl(2,R)\oplus _{s}W_{3}\}\oplus _{s}\infty A_{1}$ Lie algebra. From S.
Lie's theorem this indicates that there exists a \textquotedblleft
coordinate\textquotedblright\ transformation in which the reduced equation
is equivalent to the equation. Hence solutions of the Schr\"{o}dinger
equation are also solutions of the master equation (\ref{eq.01}). The plan
of the paper is as follows.

In Section \ref{hpz} we give the basic properties and definitions of Lie
symmetries and we study the existence of Lie symmetries for the master
equation (\ref{eq.01}). Furthermore we apply the zeroth-order invariants of
the Lie symmetries and we reduce the original equation to a one-dimensional
evolution equation. In Section \ref{equiv} we study the relationship between
the reduced equation and the Schr\"{o}dinger equation. Finally we draw our
conclusions and give an example in Section \ref{con}.

\section{Lie Point symmetries of the Master Equation}

\label{hpz}

For the convenience of the reader we present the basic properties and
definitions of Lie symmetries of differential equations.

Consider a differential equation $\Theta \left(
x^{k},u,u_{,i},u_{,ij}\right) =0,~$where $x^{k}$ are the independent
variables,$~$and $u=u\left( x^{k}\right) $ is the dependent variable. Then
the differential operator,
\begin{equation}
X=\xi ^{i}\left( x^{k},u\right) \partial _{i}+\eta \left( x^{k},u\right)
\partial _{u},  \label{go.10}
\end{equation}%
is called a Lie symmetry of $\Theta $, if there exists a function $\lambda $%
, such that $\mathcal{L}_{X^{\left[ 2\right] }}\Theta =\lambda \Theta $,
where $X^{\left[ 2\right] }$ is the second prolongation/extension of the
vector field $X$, in the space $\left\{ x^{k},u,u_{,i},u_{ij}\right\} ~$\cite%
{Olver,Bluman}.

Lie symmetries of differential equations can be used in order to determine
invariant solutions or transform solutions to solutions \cite{Bluman}. From
the Lie symmetry condition one defines the associated Lagrange's system%
\begin{equation}
\frac{dx^{i}}{\xi ^{i}}=\frac{du}{\eta }=\frac{du_{i}}{\eta _{\left[ i\right]
}}=...=\frac{du_{ij..i_{n}}}{\eta _{\left[ ij...i_{n}\right] }}
\end{equation}%
the solution of which provides the characteristic functions
\begin{equation}
\Lambda ^{\left[ 0\right] }\left( x^{k},u\right) ,~\Lambda ^{\left[ 1\right]
i}\left( x^{k},u,u_{i}\right) ,...,\Lambda ^{\left[ n\right] }\left(
x^{k},u,u_{,i},...,u_{ij...i_{n}}\right) .
\end{equation}%
The solution $\Lambda ^{\left[ k\right] }$ is called the kth-order invariant
of the Lie symmetry vector, (\ref{go.10}). These invariants can be used in
order to reduce the order or the number of the independent variables of the
differential equations. Another important feature of Lie symmetries of
differential equations is that they span the Lie algebra $G_{L}$. \ The
application of a Lie symmetry to $\Theta $ leads to a new differential
equation $\bar{\Theta}$ which is different from $\Theta $ and possibly
admits Lie symmetries which are not Lie symmetries of $\Theta $. This means
that the reduced equation can have properties different from the original
equation. However, the solutions of these equations are related through the
point transformation which transformed $\Theta $, to $\bar{\Theta}$.

\subsection{The Master Equation}

In order to simplify the presentation of the calculations we rewrite
equation (\ref{eq.01}) in the following form\footnote{%
It is possible to apply also a coordinate transformation, $\left( x,y\right)
\rightarrow \left( \bar{x},\bar{y}\right) $, which \textquotedblleft
diagonalises\textquotedblright\ the second derivatives in equation (\ref%
{eq.01}). However, we prefer to work on the original physical system.}
\begin{equation}
-\frac{x}{m}Z_{,y}+p(t)yZ_{,x}+q(t)\left( xZ\right)
_{,x}+r(t)Z_{,xx}+s(t)Z_{,xy}-Z_{,t}=0,  \label{eq.02}
\end{equation}%
where $p\left( t\right) =m\Omega ^{2}(t)$,$~q\left( t\right) =2\Gamma (t)$, $%
r\left( t\right) =%
\hbar%
m\Gamma (t)h(t)$, $s\left( t\right) =%
\hbar%
\Gamma (t)f(t)$.

We assume the generator of the one-paramete infinitesimalr point
transformation to be~%
\begin{equation}
X=\xi ^{t}\partial _{t}+\xi ^{x}\partial _{x}+\xi ^{y}\partial _{y}+\eta
\partial _{Z},  \label{eq.03}
\end{equation}%
in which~$\xi ^{t},\xi ^{x},\xi ^{y}$ and $\eta $ are functions of $\left\{
t,x,y,Z\right\} $. Furthermore, because equation (\ref{eq.02}) is a linear
equation, we have that $\eta =G\left( t,x,y\right) Z+G_{0}Z+b\left(
t,x,y\right) $, where $b\left( t,x,y\right) $ are solutions of equation (\ref%
{eq.02}) and form the infinite-dimensional Lie algebra $\infty A_{1},~$\cite%
{Bluman1}.

Hence from the Lie symmetry condition we have that\footnote{%
In this work we used the symbolic package Sym for Mathematica \cite{Dimas05a}%
.}
\begin{equation}
\xi ^{t}=a(t)~,~\xi ^{y}=f_{1}(t),  \label{eq.04}
\end{equation}%
\begin{equation}
\xi ^{x}=m\frac{-f_{1}ps+2rf_{1}^{\prime }+m\left( qsf_{1}^{\prime
}+f_{1}^{\prime }s^{\prime }-sf_{1}^{\prime \prime }\right) }{\left(
2r+m\left( 2qs+s^{\prime }\right) \right) }  \label{eq.05}
\end{equation}%
and%
\begin{eqnarray}
G &=&-\left( 2r+m\left( 2qs+s^{\prime }\right) \right) ^{-2}  \nonumber \\
&&+\;s^{\prime })+p2(x+myq)r+m2xqs+2myq^{2}s  \nonumber \\[0.01in]
&&+\;2mysq^{\prime }+2yr^{\prime }+xs^{\prime }+3myqs^{\prime }+mys^{\prime
\prime })  \nonumber \\[0.01in]
&&+\;m2m^{2}yq^{3}sf_{1}^{\prime }-2myrf_{1}^{\prime }q^{\prime
}-m^{2}yf_{1}^{\prime }q^{\prime }s^{\prime }  \nonumber \\[0.01in]
&&+\;mq^{2}f_{1}^{\prime }\left( 2yr+2xs+3mys^{\prime }\right)
-y(pf_{1}^{\prime }2r+m(2qs  \nonumber \\[0.01in]
&&+\;s^{\prime }))+2xrf_{1}^{\prime \prime }+2m^{2}ysq^{\prime
}f_{1}^{\prime \prime }+2myr^{\prime }f_{1}^{\prime \prime }+mxs^{\prime
}f_{1}^{\prime \prime }+m^{2}yf_{1}^{\prime \prime }s^{\prime \prime }
\nonumber \\[0.01in]
&&-\;2myrf_{1}^{\prime \prime \prime }-m^{2}ys^{\prime }f_{1}^{\prime \prime
\prime }+q2xrf_{1}^{\prime }+m(f_{1}^{\prime }\left( 2yr^{\prime
}+xs^{\prime }+mys^{\prime \prime }\right) )  \nonumber \\[0.01in]
&&+\;2\left( xsf_{1}^{\prime \prime }+mys^{\prime }f_{1}^{\prime \prime
}-mysf_{1}^{\prime \prime \prime }\right) )))),  \label{eq.06}
\end{eqnarray}%
where $G_{0}$ is a constant and prime means differentiation with respect to
time, \textquotedblleft $t$\textquotedblright $,$ and functions $a\left(
t\right) $ and $f_{1}\left( t\right) $ are related to $p,q,r,s$ by a system
of ordinary differential equations which me omit. We can see that $%
f_{1}\left( t\right) $ satisfies a linear fourth-order differential equation
which means that it provides us with four symmetries. Another symmetry
vector arises from the unique solution of $a\left( t\right) $. Therefore
from (\ref{eq.04})-(\ref{eq.06}) it is easy to see that the Lie symmetries
of the master equation form the $\{A_{1}\oplus _{s}W_{5}\}\oplus _{s}\infty
A_{1}$ Lie algebra.

Indeed the form of the symmetry vector (\ref{eq.03}) it is not a closed
form. The reason for this is that we have considered arbitrary functions, $%
\Omega ^{2}\left( t\right) $,~$\Gamma \left( t\right) ,~h\left( t\right) $
and $f\left( t\right) .$ In a case for specific functional forms of the
coefficients one can calculate the symmetry vector in closed-form. For
instance in the case for which the coefficients, $p,q,r~$and $s, $ are
constants the Lie symmetries are
\begin{equation}
Y_{1}=a_{1}\partial _{t}~,~Y_{Z}=Z\partial _{Z}~,~Y_{b}=b\partial _{z},
\label{eq.06a}
\end{equation}%
\begin{equation}
X_{1}=e^{\frac{\lambda -q}{2}t}\left[ m\left( \lambda -q\right) \partial
_{x}+2\partial _{y}\right] ~,~X_{2}=e^{-\frac{\lambda +q}{2}t}\left[ m\left(
\lambda +q\right) \partial _{x}-2\partial _{y}\right] ,  \label{eq.06b}
\end{equation}%
\begin{equation}
X_{3}=e^{-\frac{\left( \lambda -q\right) }{2}t}\left[ 2rm\left( q-\lambda
\right) \partial _{x}+4\left( r+sqm\right) \partial _{y}+\left( 2mq\left(
\lambda -q\right) x+m^{2}q\left( \lambda ^{2}-q^{2}\right) y\right)
Z\partial _{Z}\right]  \label{eq.06c}
\end{equation}%
and
\begin{equation}
X_{4}=e^{\frac{\lambda +q}{2}t}\left[ 2rm\left( \lambda +q\right) \partial
_{x}+\left( r+sqm\right) \partial _{y}+\left( -2mq\left( \lambda +q\right)
x+m^{2}q\left( \lambda ^{2}-q^{2}\right) y\right) Z\partial _{Z}\right] ,
\label{eq.06d}
\end{equation}%
where $\lambda =\sqrt{4p-mq^{2}}$.

Below we apply the zeroth-order invariants of the Lie symmetry which
corresponds to the solution of the function $f_{1}\left( t\right) $.

\subsection{Application of the Lie invariants}

Consider now the Lie symmetry vector (\ref{eq.03}) for which $G_{0}=0$ and $%
a\left( t\right) =0$. The characteristic functions are%
\begin{equation}
Z\left( t,x,y\right) =U(t,B(t)f_{1}(t)x-A(t)y)\exp [J(t,x)\left(
B(t)f_{1}(t)x-A(t)y\right) ],  \label{eq.07}
\end{equation}%
where
\begin{equation}
J=\frac{x}{2A^{2}}\left[
\begin{array}{c}
A^{2}\left( 2G+xH\right) + \\
~+B\left( -2\left( Bf_{1}x-A(t)y\right) +xBf_{1}\right) K%
\end{array}%
\right]  \label{eq.08}
\end{equation}%
and the functions, $A\left( t\right) ,B\left( t\right) ,K\left( t\right)
,G\left( t\right) $ and $H\left( t\right) $, are the coefficients of the
symmetry vector (\ref{eq.03}).

Hence the application of (\ref{eq.07}) to (\ref{eq.02}) gives the reduced
equation
\begin{equation}
S\left( t\right) U_{,ww}-wR\left( t\right) U_{,w}+q\left( t\right)
U-U_{,t}=0,  \label{eq.09}
\end{equation}%
where $w=B(t)f_{1}(t)x-A(t)y,$~$S\left( t\right) =Bf_{1}\left(
Bf_{1}r-As\right) ~$and$~R\left( t\right) =A^{-1}\left( Bf_{1}p-A^{\prime
}\right) \,$.

This means that the Lie symmetries provide us with a solution for the master
equation (\ref{eq.01}), which is (\ref{eq.07}) and the function $U\left(
t,w\right) $ is given by (\ref{eq.09}). Below we study the Lie symmetries of
(\ref{eq.09}) and we show that it is maximally symmetric. This means that it
is equivalent with the elementary one-dimensional Schr\"{o}dinger Equation.

\section{Equivalence with the Schr\"{o}dinger Equation}

\label{equiv}

For simplicity in the following we consider a \textquotedblleft
time\textquotedblright\ rescaling, $t\rightarrow T$, such that $S\left(
T\right) =1$. \ Hence equation (\ref{eq.09}) becomes%
\begin{equation}
U_{,ww}-wR\left( T\right) U_{,w}+q\left( T\right) U-U_{,T}=0.  \label{eq.10}
\end{equation}

We apply the Lie symmetry condition to the this equation and we derive the
following symmetry vector field%
\begin{equation}
Y=\alpha (T)\partial _{t}+\left[ \frac{\dot{\alpha}(T)v}{2}+\beta (T)\right]
\partial _{v}+\left( F(T,w)U+\bar{b}\left( t,w\right) \right) \partial _{U},
\label{eq.11}
\end{equation}%
where
\begin{equation}
F(T,v)=\phi (T)+\frac{1}{4}\left[ 2v\beta (T)R(T)+v^{2}R(T)\dot{\alpha}-2v%
\dot{\beta}+v^{2}\alpha (T)\dot{R}-\frac{1}{2}v^{2}\ddot{\alpha}\right] ,
\end{equation}%
in which overdot denotes differentiation with respect to $T$ and the
functions $\phi (T)$, $\beta (T)$ and $\alpha (T)$ are solutions of the
equations
\begin{eqnarray}
\phi &=&\phi _{0}+\alpha (\left( q+\frac{1}{2}R\right) -\frac{1}{4}\dot{%
\alpha},  \label{eq.12} \\
\ddot{\beta} &=&\left( \dot{R}+R^{2}\right) \beta \quad \mbox{\rm and}
\label{eq.13} \\
\mathinner{\buildrel\vbox{\kern5pt\hbox{...}}\over{\alpha}} &=&4\dot{\alpha}%
\left( \dot{R}+R^{2}\right) +2\alpha \frac{d}{dT}\left( \dot{R}+R^{2}\right)
.  \label{eq.14}
\end{eqnarray}%
Furthermore $\bar{b}\left( t,w\right) $ satisfies the original equation, (%
\ref{eq.10}).

Equation (\ref{eq.13}) is a maximally symmetric linear second-order
differential equation. In this case, by application of the Riccati
transformation $R=\frac{\dot{L}}{L}$,~in (\ref{eq.13}) we find the solution,%
\begin{equation}
\beta \left( T\right) =\beta _{0}L\left( T\right) +\beta _{1}L\left(
T\right) \int L^{-2}\left( T\right) dT.
\end{equation}

Equation (\ref{eq.14}) is a nonautonomous third-order differential equation.
We multiply with $\alpha \left( T\right) $ and integrate to obtain%
\[
\alpha (T)\ddot{\alpha}-\frac{1}{2}\dot{\alpha}^{2}-2\alpha ^{2}(T)\left(
\dot{R}+R^{2}(T)\right) =2K,
\]%
where $K$ is a constant. We substitute $\alpha =\gamma ^{2}$ into this
equation and hence we find the well-known Ermakov-Pinney equation \cite%
{Erm1,Erm2}%
\begin{equation}
\ddot{\rho}-\rho (T)\left( \dot{R}+R^{2}(T)\right) =\frac{K}{\rho ^{3}(T)}.
\label{eq.15}
\end{equation}%
The solution of (\ref{eq.15}) is given in \cite{Erm2} and it is related with
the solution of the linear equation%
\begin{equation}
\ddot{\sigma}-\left( \dot{R}+R^{2}(T)\right) \sigma =0.
\end{equation}

Therefore we conclude that equation (\ref{eq.10}) admits as Lie symmetries
the vector fields which form the $\{sl(2,R)\oplus _{s}W_{3}\}\oplus
_{s}\infty A_{1}$ Lie algebra. Hence from S. Lie's theorem \cite{lie} we
have that there exists a transformation, $\left( T,w,U\right) \rightarrow
\left( \tau ,\chi ,\Psi \right) ,$ in which (\ref{eq.10}) becomes%
\begin{equation}
-\frac{%
\hbar%
}{2M}\frac{\partial ^{2}\Psi }{\partial \chi ^{2}}=i%
\hbar^{2}%
\frac{\partial \Psi }{\partial \tau }  \label{eq.16}
\end{equation}%
which is the Schr\"{o}dinger equation for a free particle. That is possible
because equations (\ref{eq.10}) and (\ref{eq.16}) are both maximally
symmetric.

\section{Discussion}

\label{con}

In this work with the application of the group invariant transformations we
proved that there exists a surface in the space of the dependent and
independent variables in which the master equation (\ref{eq.01}) can be seen
as a one-dimensional equation. That means that solutions of the latter
generate solutions for the master equation given by the expression (\ref%
{eq.07}), that is, there is class of solutions which describe the
two-different systems, but the solutions are given in different
representations.

We remark that in our analysis we considered that the coefficients of the
master equation are arbitrary functions of time, which means that the result
holds when the coefficients are constants. For instance, consider the
application of the Lie symmetry $X_{3}$,~(\ref{eq.06b}), in equation (\ref%
{eq.02}) for constant coefficients. Hence we have that $Z=U\left( t,w\right)
,~$where~$w=\frac{ym\left( L-q\right) -2x}{m\left( L-q\right) }$, and $%
U\left( t,w\right) $ satisfies the equation%
\begin{equation}
\bar{s}U_{,ww}-\frac{\left( \lambda +q\right) }{2}wU_{,w}+2qU-U_{,t}=0
\label{eq.17}
\end{equation}%
and $\bar{s}=2\frac{-2r+sm\left( \lambda -q\right) }{m^{2}\left( \lambda
-q\right) ^{2}}$. Therefore under the coordinate transformation,
\begin{equation}
U\left( t,w\right) =e^{2qt}\Psi \left( \tau ,\chi \right) ~,~w=\left( \frac{%
2Ms}{%
\hbar%
}\right) ^{1/2}\chi e^{\left( \frac{\lambda +q}{2}t\right) }~,~d\tau =-i%
\hbar%
e^{-\left( \lambda +q\right) t}dt,
\end{equation}%
the latter equation takes the form of the one-dimensional Schr\"{o}dinger
equation. A similar result holds and for the remaining Lie symmetry vectors,
$X_{2}-X_{4}$, or any linear combination of them.

\subsubsection*{Acknowledgements}
The research of AP was supported by FONDECYT postdoctoral grant no. 3160121.
AP thanks the Durban University of Technology for the hospitality provided
while part of this work was performed. RMM thanks the National Research
Foundation of the Republic of South Africa for the granting of a
postdoctoral fellowship with grant number 93183 while this work was being
undertaken. RS thanks the University Grants Commission of India for support.

\end{document}